\documentclass[12pt]{revtex4}
\usepackage{amsmath,amssymb,bm}
\usepackage[toc,page]{appendix}
\usepackage{graphicx}
\usepackage{url}
\usepackage{hyperref,xcolor}

\DeclareGraphicsExtensions{.pdf,.png,.jpg,.eps}

\begin{document}
\title{Dynamical inhomogeneity of amorphous solid and its geometric features}
\author{Cunyuan Jiang}
\email{cunyuanjiang@zzu.edu.cn}
\affiliation{School of Physics, Zhengzhou University, Zhengzhou 450001, Henan, China}

\begin{abstract}
     The dynamical response of amorphous solids is expected to be inhomogeneous due to the absence of translational symmetry, unlike in crystals. In this work, we use Green's function method combined with the Hessian matrix from simulation data to characterize the vibrational inhomogeneity of a two-dimensional amorphous solid. We define an order parameter for vibrational inhomogeneity, the frequency-resolved participation ratio, which equals a constant value of one over all frequencies for a crystal, while being around \(\sim 0.7\) for almost the entire frequency range in an amorphous solid, thus indicating a similar level of vibrational inhomogeneity across most of the frequency range. The spatial distribution of vibrational inhomogeneity in the amorphous solid is presented in three ways: by frequency, at the particle level, and via percolation analysis, all showing that the vibrational inhomogeneity is fragmented and occurs at the particle scale. The results suggest that the Green's function method is a convenient tool for probing vibrational inhomogeneity, and provide an alternative perspective on the role of vibrational inhomogeneity in amorphous solids.
\end{abstract}
\maketitle

\section{Introduction}
The translational symmetry of crystalline materials ensures that all particles contribute equally to vibrations (when each unit cell contains only one particle) at equilibrium\cite{Kittel2004}. For amorphous solids, such translational symmetry is absent and hence particles may not contribute equally to vibrations, depending on their local environments. Vibrational inhomogeneity in amorphous solids has attracted great interest due to its association with low-frequency local dynamical anomalies such as localized modes\cite{Wang2019,PhysRevLett.108.095501,10.1063/5.0147889}, soft spots\cite{Corrado2020}, dynamical defects\cite{ch39-6bhs,PhysRevResearch.6.023053}, and dynamical excitations\cite{Hu2022,PhysRevLett.133.188302}. These local dynamical anomalies are also considered to be correlated with global dynamical anomalies such as the boson peak (BP)\cite{Hu2022,PhysRevLett.133.188302,ch39-6bhs}.

In the literature, the identification of vibrational inhomogeneity is mostly based on eigenmode analysis of the dynamical matrix. For example, vibrational inhomogeneity can be seen when the displacement vector of a given particle is larger or smaller than those of others in an eigenmode. An order parameter, the participation ratio, can be applied directly to an eigenmode \(\phi_\alpha(i)\) to estimate inhomogeneity\cite{PhysRevLett.125.085502,PhysRevLett.117.035501,10.1063/5.0147889,Wang2019}, through the following definition,
\begin{equation}
    \text{PR}_\alpha = \frac{\left( \sum_{i=1} |\phi_\alpha(i)|^2 \right)^2}{2N \sum_{i=1} |\phi_\alpha(i)|^4} . \label{preigen}
\end{equation}
Here \(N\) is the total number of particles, \(\alpha\) denotes the index of eigenmode, and \(i\) denotes the index of degree of freedom. The above definition of the participation ratio faces two issues that make it not a perfect order parameter for characterizing vibrational inhomogeneity. The first issue is the ambiguity of the physical meaning of an eigenmode. For example, as mentioned earlier, the dynamical response in a crystal should be homogeneous due to symmetry, yet the eigenmodes are plane waves with spatial fluctuations in intensity. Therefore, the participation ratio of a crystal's eigenmode is not constantly \(1\), but a fractional number. This point will be discussed in more detail in the following sections. More importantly, the spatial distribution of the dynamical response cannot be obtained correctly from a single eigenmode due to intensity fluctuations. The second issue is the sensitivity of the definition to numerical values. If \(1\%\) of particles have a strength of \(10\) and \(99\%\) have strength \(1\), the participation ratio is \(0.6\); however, this value drops sharply to \(0.04\) if the \(1\%\) have strength \(100\) instead of \(10\). This sensitivity makes the order parameter a poor estimator of how many particles contribute to the vibration. Hence, the spatial distribution is still required to fully understand vibrational inhomogeneity in amorphous solids.

The Green's function method is more commonly used in theoretical analyses of amorphous systems\cite{PhysRevLett.100.137402,PhysRevLett.98.025501,Ding2025}, because it conveniently characterizes the dynamical response under external excitations. Mathematically, the Green's function is the solution of the dynamical equation with a homogeneous pulse excitation applied to every particle\cite{Jiang_2026}\footnote{Here, homogeneous excitation on every particle results from the unit matrix \(I\) on the right-hand side of Eq.\eqref{eqdynamics}, for which the diagonal elements are constantly \(1\). The diagonal element \(G_{ii}\) of the Green's function matrix gives the response at degree of freedom \(i\) to a source applied at the same DOF.},
\begin{equation}
    \left( \partial_t^2 - H \right) G (t - t')= I \delta (t - t'). \label{eqdynamics}
\end{equation}
Here, \(H\) is the dynamical matrix, \(G(t-t')\) is the Green's function in the time domain, and \(I\) is the unit matrix. Therefore, the diagonal element \(G_{ii}(t-t')\) of the Green's function gives the local dynamical response of the \(i\)th degree of freedom under homogeneous pulse excitation. For amorphous solids, as mentioned earlier, the local dynamical response may differ from particle to particle due to variations in local environments, even though the excitation is the same. The Green's function method thus provides a tool to directly probe vibrational inhomogeneity with a clear physical meaning.

In this work, we use the Green's function method to probe the vibrational inhomogeneity of a two-dimensional amorphous solid from simulations. We define the total vibrational density of states (DOS), the local DOS (which represents frequency-resolved vibrational inhomogeneity), and the frequency-resolved participation ratio \(\text{PR}(\omega)\) to characterize the dynamical properties of the amorphous solid. We find that the vibrational inhomogeneity remains at a similar level, \(\text{PR}(\omega) \sim 0.7\), over most of the frequency range. The spatial features of vibrational inhomogeneity are fragmentation and occur at the particle scale. The findings call for noticing inhomogeneity in amorphous solids from broader perspectives of dynamics and structure, and across different frequency and time scales.

\section{Simulation and theory}
To obtain an amorphous solid structure from simulations, we annealed \(4096\) bidisperse particles with a 1:1 mixture ratio and a size ratio of 1.4:1, interacting via the harmonic potential \(V(r_{ij})=\frac{\epsilon}{2}(1-r_{ij}/\sigma_{ij})^2\) for \(r_{ij}<\sigma_{ij}\) (zero otherwise) in two dimensions (2D). All particles have equal mass \(m\), and periodic boundary conditions are applied with a fixed area fraction \(\phi=0.91\) in the canonical ensemble. (Length, mass, energy, and time are expressed in reduced Lennard-Jones units: the small particle diameter \(\sigma_{\rm s}\), \(m\), \(\epsilon\), and \(\sqrt{m\sigma_{\rm s}^2/\epsilon}\), respectively, with \(k_{\rm B}=1\).) The system is first equilibrated at \(T=3.0\) for \(10\,000\) time units, then cooled stepwise with \(\Delta T=0.1\) down to \(T=0.1\), with an equilibration time \(\Delta t\) at each step that sets the cooling rate \(\gamma=\Delta T/\Delta t\) (ranging from \(10^{-3}\) to \(10^{-6}\)). Further simulation details can be found in Ref.\cite{Tong2020}.
 
At the configuration with minimum total potential energy \(U\), the Hessian matrix can be obtained by computing the second derivatives with respect to each degree of freedom (DOFs) \(x_{i(j)}\): \(H_{ij} = \partial^2 U / \partial x_i \partial x_j\). Here, the DOFs \(x_{i(j)}\) include both the \(x\) and \(y\) coordinates of all particles. The dynamical response properties can be obtained using the Green's function (\(G\)) method once the Hessian matrix \(H\) of the system is known. Their relation in the frequency domain can be directly obtained upon Fourier transform of Eq.\eqref{eqdynamics}, as,
\begin{equation}
    G(\omega) = [(\omega^2 + i\eta) I - H]^{-1}, \label{greenfunction}
\end{equation}
with \(\eta = 0^+\) a small number ensuring causality, \([]^{-1}\) denoting the inverse of a matrix, and \(I\) the identity matrix. The total DOS can be easily obtained as\cite{PhysRevLett.122.145501},
\begin{equation}
    g(\omega) = -\omega \mathrm{Im}[\mathrm{Tr}[G(\omega)]] \label{dos}
\end{equation}
with \(\mathrm{Tr}[]\) computing the trace of a matrix. The local DOS on the \(i\)th DOF is simply,
\begin{equation}
    g_i(\omega) = -\omega \mathrm{Im}[G_{ii}(\omega)] . \label{localdos}
\end{equation}
For convenience in displaying BP, the DOS curve is usually plotted after reduction by Debye's law: \(g(\omega)/\omega = -\mathrm{Im}[\mathrm{Tr}[G(\omega)]]\) and \(g_i(\omega)/\omega = -\mathrm{Im}[G_{ii}(\omega)]\). 

The local DOS defined in Eq.\eqref{localdos} provides a direct measure of how much each particle contributes to the total DOS at a given frequency, as guaranteed by the identity
\begin{equation}
    g(\omega) \equiv \sum_i g_i (\omega).
\end{equation}
The advantage of the local DOS is that it reveals the spatial distribution of the dynamical response at a specific frequency. In our opinion, however, this approach is relatively limited for characterizing local dynamical objects such as localized modes\cite{Wang2019,PhysRevLett.108.095501,10.1063/5.0147889}, soft spots\cite{Corrado2020}, and even topological defects\cite{PhysRevMaterials.4.113609} mentioned earlier.

To identify the morphological properties of the distribution of local DOS at a given frequency, \(g_i (\omega)\), we define frequency resolved participation ratio \(\text{PR} (\omega)\) as,\cite{Hu2022}
\begin{equation}
    \text{PR} (\omega) = \dfrac{(\sum_i g_i(\omega))^2}{2 N \sum_i g_i (\omega)^2} . \label{frpr}
\end{equation}
Frequency resolved participation ratio \(\text{PR} (\omega) = 1\) indicates all particles contribute to the mode at frequency \(\omega\) and hence the mode is extended and dynamically homogeneous, while \(\text{PR} (\omega) = 1/N\) indicates only one particle contributes, and hence the mode is localized and strongly dynamically inhomogeneous. The physics captured by the frequency-resolved participation ratio in Eq.\eqref{frpr} differs from that of the eigenmode participation ratio in Eq.\eqref{preigen}. The frequency-resolved PR characterizes the distribution of vibrational intensity, i.e., vibrational inhomogeneity, at a given frequency, whereas the eigenmode PR characterizes the vibrational inhomogeneity of a single eigenmode. More generally, \(\text{PR}(\omega)\) cannot be obtained simply from the eigenmode PR (\(\text{PR}_\alpha\)):
\begin{equation}
    \text{PR} (\omega) \neq \sum_\alpha \delta(\omega - \omega_\alpha) \text{PR}_\alpha . \label{inequality}
\end{equation}
This inequality is clearly seen in the bottom panels of Fig.\ref{figdosandpr}. At a given frequency, there can be many eigenmodes with nearly degenerate eigenfrequencies. Their relative contributions to the vibration at that frequency are, in principle, different. However, the normalization in Eq.\eqref{preigen} discards the weighting information for different eigenmodes. On the other hand, the frequency-resolved PR does not rely on any eigenmode decomposition, and is therefore able to correctly characterize the inhomogeneity of the vibrational response at a given frequency.

\section{Results and discussion}

\subsection{Comparison of results from the Green's function and eigenmode analysis}

\begin{figure}[htbp]
    \centering
    \includegraphics[width=0.8\linewidth]{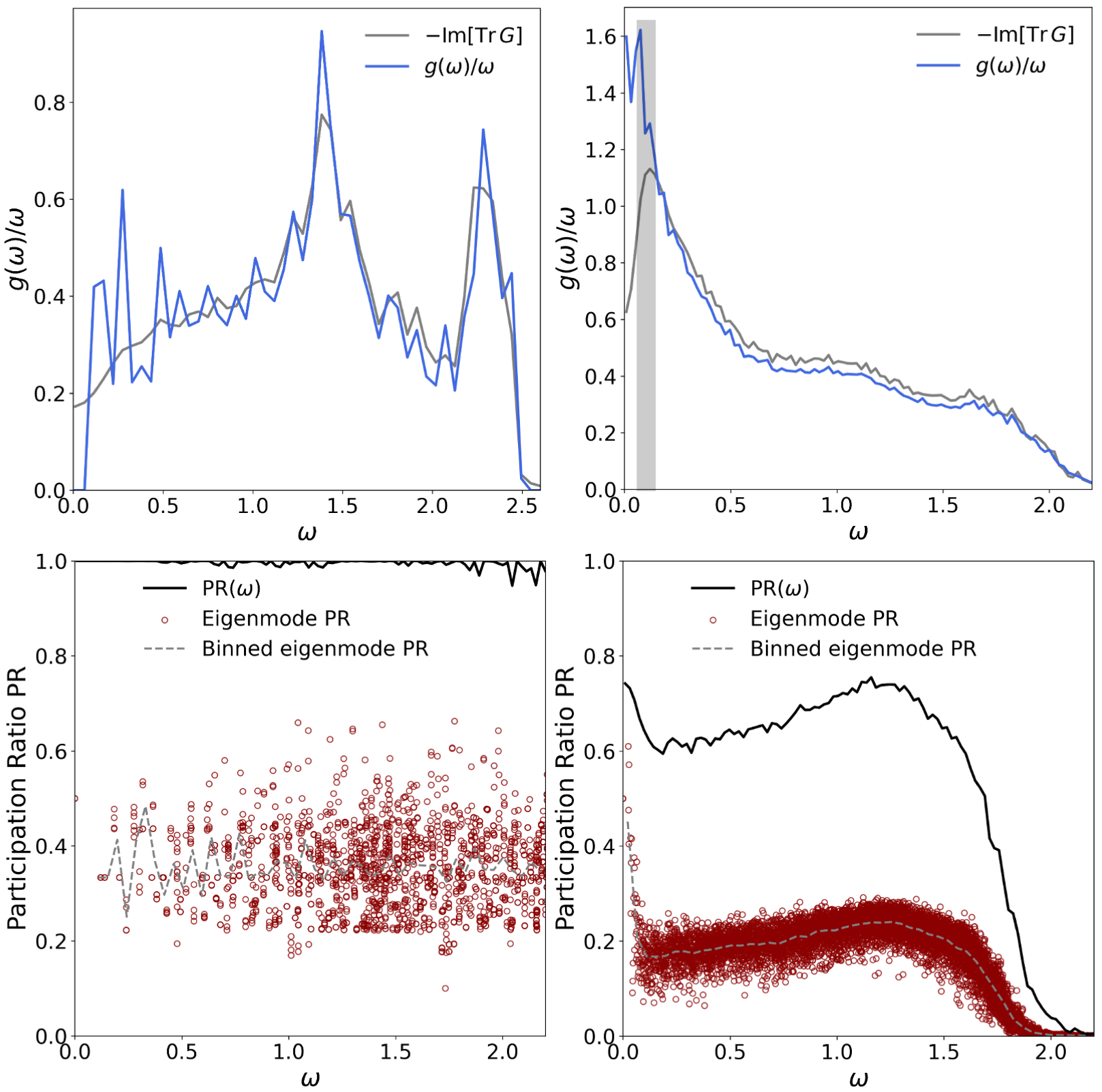}
    \caption{Panels \textbf{top left} (for crystal case) and \textbf{top right} (for amorphous solid case) show the reduced DOS \(-\mathrm{Im}[\mathrm{Tr}[G(\omega)]]\) obtained from Green's function method (\textbf{gray}), and \(g(\omega) / \omega = \sum_\alpha \delta(\omega - \omega_\alpha) / \omega \) obtained from the distribution of eigenfrequencies of Hissian matrix (\textbf{gray}). In the top right panel, frequency of BP is identified to be around \(\omega_{BP} \sim 0.15\). The \textbf{bottom} two panels show the participation ratio of crystal (\textbf{left}) and amorphous solid (\textbf{right}) cases obtained with the method of Eq.\eqref{preigen} (the red dots) and Eq.\eqref{frpr} (the black line). The gray dashed lines in bottom panels are frequency binned participation ratio of eigenmodes.}
    \label{figdosandpr}
\end{figure}

Low-frequency sound waves are collective excitations in crystals. For each sound wave mode, all particles participate equally because of translational symmetry. Therefore, the DOS of a crystal should follow Debye's law\cite{debye} in the low-frequency region, and the frequency-resolved participation ratio should be \(1\) over the entire frequency range. In the top left panel of Fig.\ref{figdosandpr}, we validate the Green's function method for a hexagonal close-packed crystal by comparing the reduced total DOS with the distribution of eigenfrequencies. The two reduced DOS curves coincide in trend over the entire frequency range. The Green's function result is smoother than the eigenfrequency statistics due to the finite broadening parameter \(\eta\) in Eq.\eqref{greenfunction}. Theoretically, a smaller \(\eta\) brings the result closer to the eigenfrequency distribution, while a larger \(\eta\) yields a smoother curve. To balance accuracy and visual clarity, we use \(\eta \sim 10^{-2}\) in our computations. Appendix A shows the effect of \(\eta\) on the reduced DOS curves in more detail. In the bottom left panel of Fig.\ref{figdosandpr}, we compare the two participation ratios obtained via Eq.\ref{preigen} (\(\text{PR}_\alpha\)) and Eq.\ref{frpr} (\(\text{PR}(\omega)\)). As discussed earlier, the frequency-resolved PR remains constantly \(1\) over all frequencies, whereas the eigenmode PR is distributed over a wide range from \(0.1\) to \(0.7\). Given the translational symmetry that ensures equal participation of all particles in a vibration at a given frequency, the frequency-resolved PR correctly characterizes the fraction of participating particles. In contrast, the eigenmode PR describes the inhomogeneity of individual eigenmodes but cannot reliably indicate the participation ratio for a vibration at a specific frequency, because of its fractional value and the inequality Eq.\eqref{inequality}.

By comparing the reduced DOS and participation ratio obtained from the Green's function method and eigenmode analysis for the crystal, we validate the Green's function approach and highlight the advantages of the participation ratio defined via Green's function for characterizing vibrational inhomogeneity, i.e., the fraction of participating particles. The top right and bottom right panels of Fig.\ref{figdosandpr} show the reduced DOS and participation ratio for the amorphous solid. For the amorphous solid, the reduced DOS from the two methods coincide at frequencies above \(\omega > 0.15\), but deviate noticeably near and below the BP (\(\omega \sim 0.15\)). Near zero frequency, the eigenfrequency statistics are usually noisy due to the sparsity of modes. A common remedy is to perform an ensemble average over many independent configurations\cite{Xi2026}. Here, we note another minor advantage of the Green's function method: the reduced DOS curve obtained without ensemble averaging is already as smooth as the averaged results over nearly forty configurations, comparable to the reduced DOS of 2D or 3D glasses in Ref.\cite{Xi2026}.

For the amorphous solid, the total reduced DOS clearly exhibits a BP at \(\omega_{BP} = 0.15\), well below the Van Hove peak (located at around \(\omega_{VH} \sim 1.5\) based on the crystal result in Fig.\ref{figdosandpr}). The frequency-resolved participation ratio displays several noteworthy features. First, it shows a dip at the BP frequency. Below the BP, \(\text{PR}(\omega) \sim 0.8\). At frequencies above the BP but below the Van Hove peak, \(\text{PR}(\omega)\) increases gradually from \(\sim 0.7\) to \(\sim 0.8\), indicating that a large fraction of particles can be excited and participate in vibrations in this range. The value \(\sim 0.7\), which is lower than the ideal value of 1 for sound waves in crystals, signals the presence of spatial vibrational inhomogeneity (see Fig.\ref{figldosnew}). Above the Van Hove peak, the frequency-resolved PR drops to zero, showing that collective dynamical responses are suppressed. The eigenmode PR follows a similar trend but takes much lower values, as in the crystal case, because it reflects only the inhomogeneity of individual eigenmodes rather than the local DOS.

\subsection{The distribution of local DOS and vibrational inhomogeneity}

\begin{figure}[htbp]
    \centering
    \includegraphics[width=0.6\linewidth]{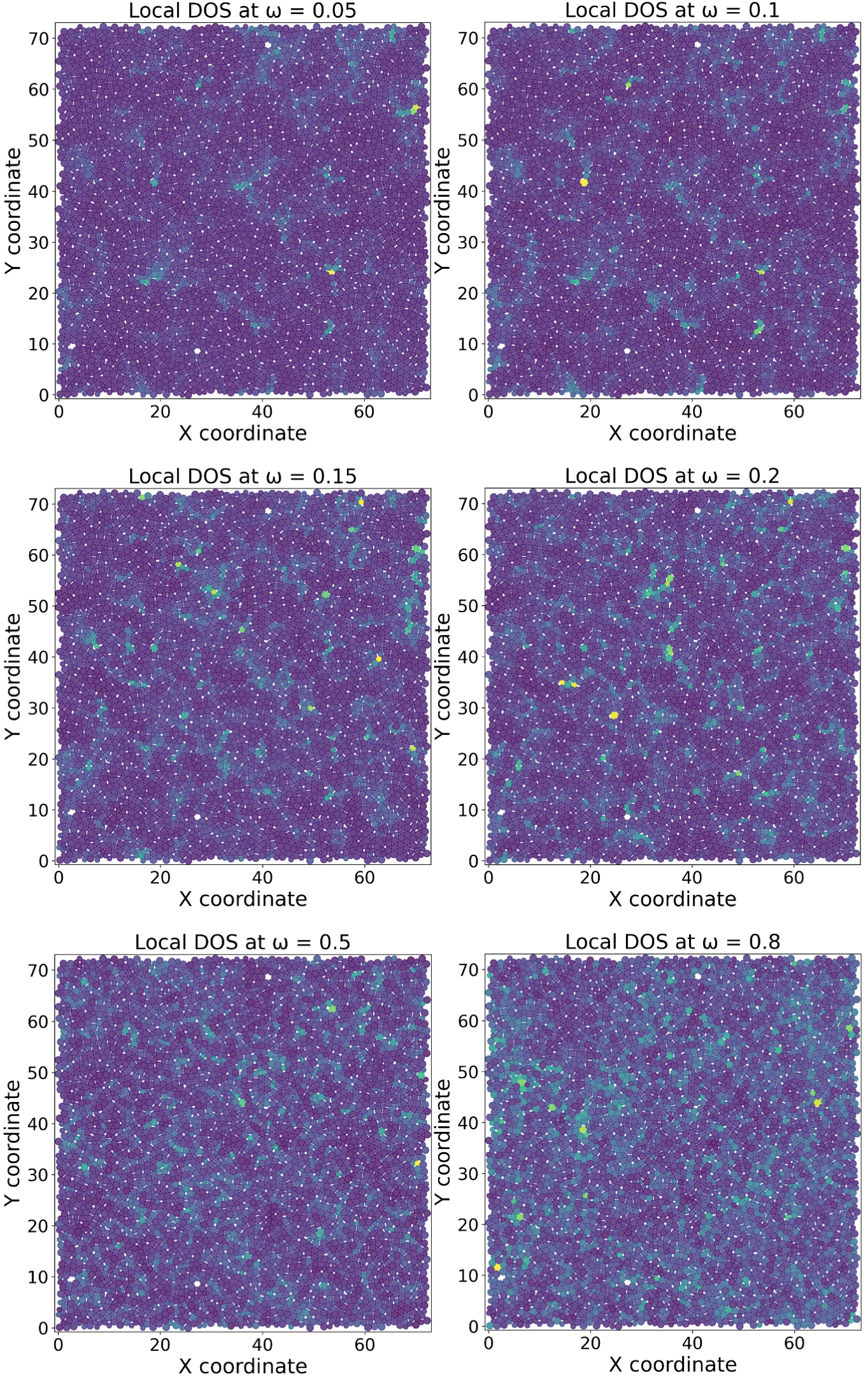}
    \caption{The spatial distribution of local DOS under frequencies, \(\omega = 0.05\), \(0.1\), \(0.15\), \(0.2\), \(0.5\), \(0.8\), from \textbf{top left} to \textbf{bottom right}. The color from dark to light indicates stronger local DOS.}
    \label{figldosnew}
\end{figure}

\begin{figure}[htbp]
    \centering
    \includegraphics[width=0.6\linewidth]{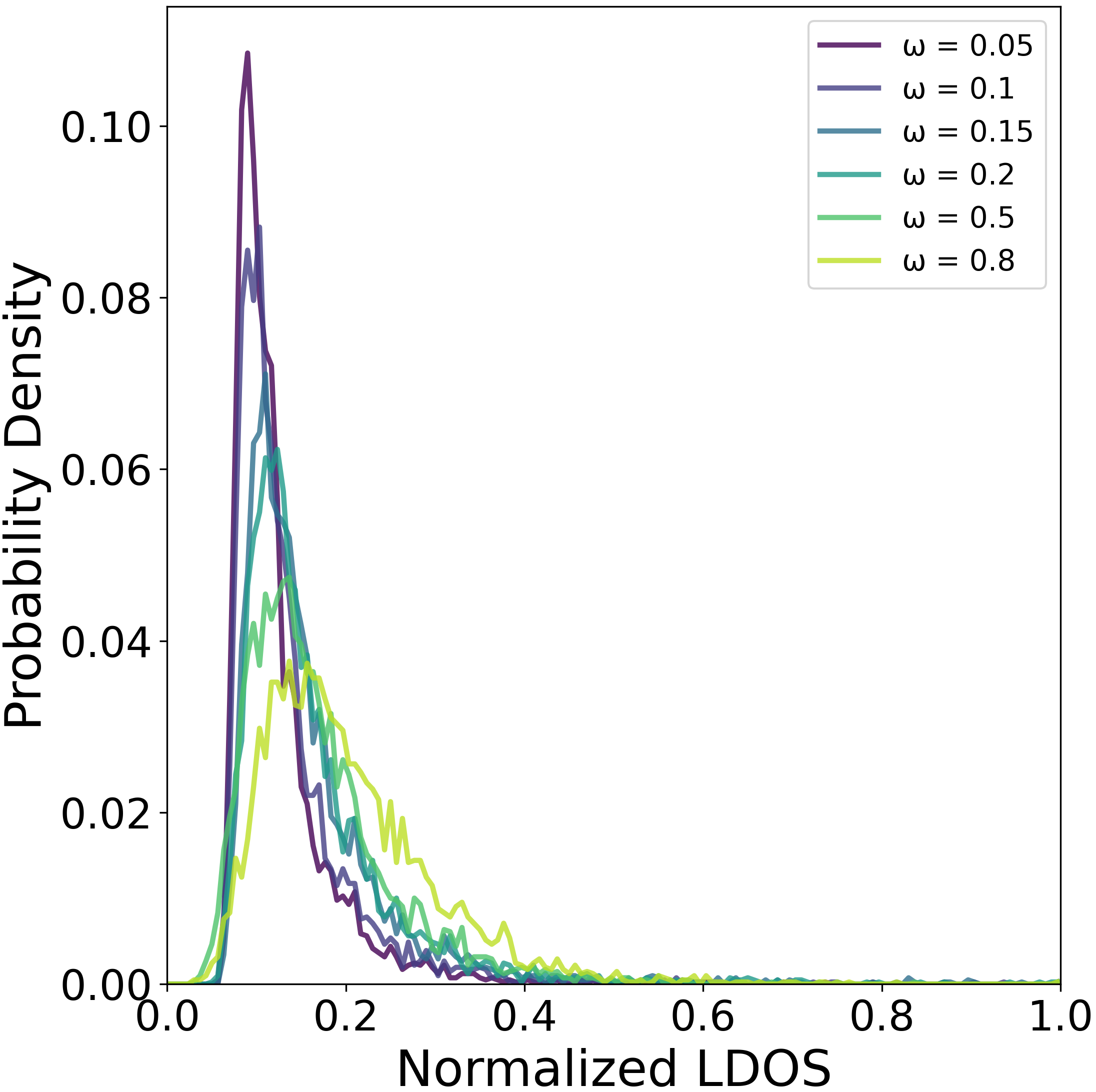}
    \caption{The distribution of local DOS along the values for different frequencies. The value of local DOS has been normalized by the maxima of each spatial distribution of local DOS. }
    \label{figldosmore}
\end{figure}

Figure \ref{figldosnew} shows that the spatial distribution of the dynamical response is fragmented and occurs at the particle scale over a broad frequency range. There are always a few particles, on the order of \(0.1\%\), exhibiting a stronger dynamical response than the rest. To give a comprehensive view of the vibrational inhomogeneity, we present the distribution of local DOS values at different frequencies in Fig.\ref{figldosmore}. The distribution is remarkably similar across a wide frequency range, from \(\omega = 0\) to \(\omega = 0.8\). The majority of local DOS values are around \(0.1\) of the maximum, while particles with very strong local DOS are extremely rare (less than \(0.1\%\)). Thus, the vast majority of particles dominate the total DOS, including the BP. The minority, despite dominating the vibrational inhomogeneity, are too few to significantly affect the total DOS.

Vibrational inhomogeneity is characterized by the non-uniform spatial distribution of the local DOS. Figure \ref{figldosmore} provides more detailed information than the frequency-resolved PR, which is an order parameter summarizing the distribution. Figure \ref{figldosmore} suggests that the distribution of local DOS shown in Fig.\ref{figldosnew} is rather uniform, and that the particles with large local DOS (light dots) in Fig.\ref{figldosnew} are a small minority.

\subsection{The BP involves the majority of particles}

\begin{figure*}[htbp]
    \centering
    \includegraphics[width=\linewidth]{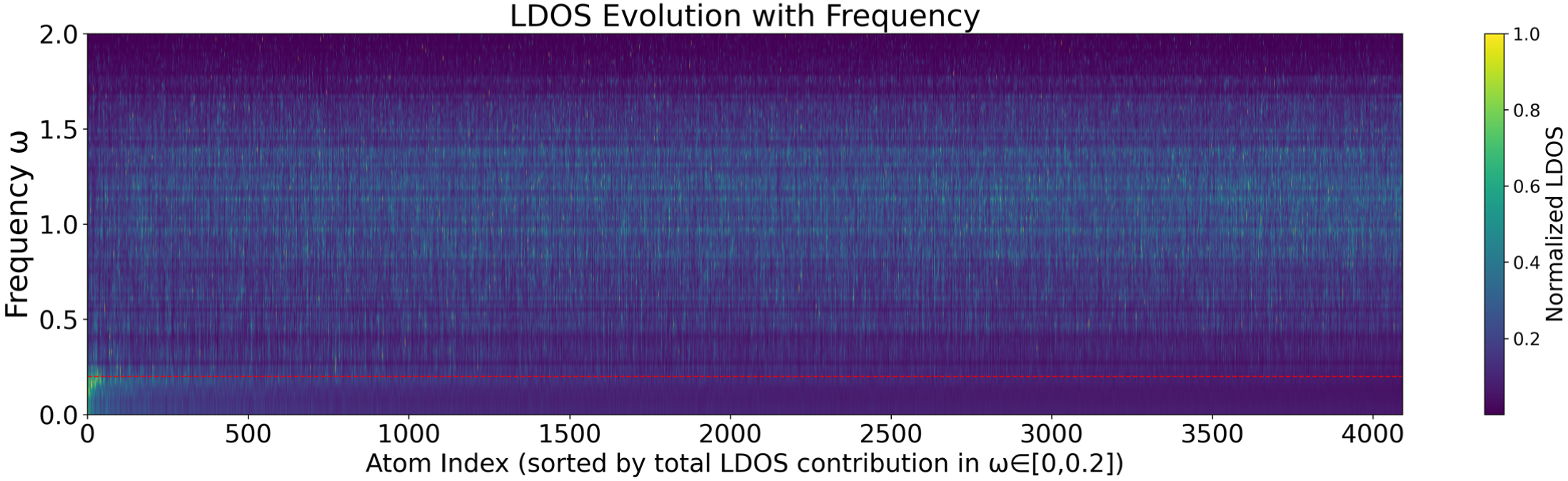}
    \caption{The local DOS as a function of frequency for all particles. The color from dark to light indicates stronger local DOS. The index of particles has been arranged according to the low frequency (\(\omega < 0.2\)) intensity, as in Eq.\eqref{ordering}.}
    \label{figevolution}
\end{figure*}

To visualize the dynamical response of all particles over the entire frequency range, we provide a heat map of the local DOS in frequency vs. all \(4096\) particles in Fig.\ref{figevolution}. The inhomogeneity can be seen from two dimensions: for a given frequency or for a single particle, the local DOS is inhomogeneous. In Fig.\ref{figevolution}, the particle indices have been sorted according to their low-frequency (\(\omega < 0.2\), around the BP) intensity. As can be seen in the lower left corner, there are indeed some particles, on the order of hundreds, that contribute ``more'' to the low-frequency total DOS.

\begin{figure}[htbp]
    \centering
    \includegraphics[width=0.8\linewidth]{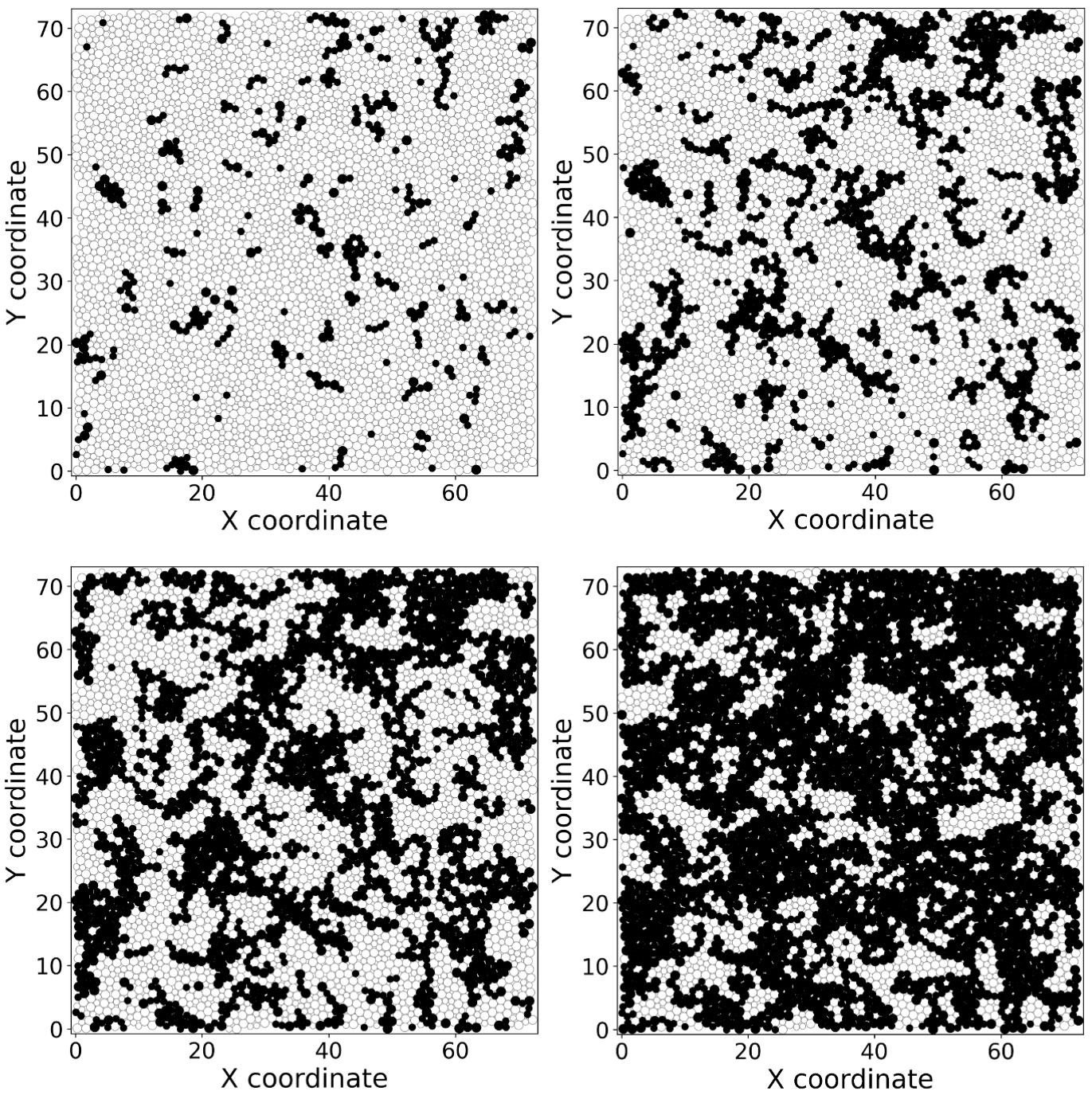}
    \caption{Percolation of the particles contributing the most to the low frequency \(\omega \in [0,0.2]\) (around BP) of total DOS. The percolation threshold, from \textbf{top left} to \textbf{bottom right}, is top \(n_{\text{DOS}} = 20\%\), \(40\%\), \(60\%\), \(80\%\) contribution of the total DOS at frequency region \(\omega < 0.2\). The ratio of survivors (colored in black) under percolation are \(n_{\text{particle}} = 9.2 \%\), \(24.5 \%\), \(44.6 \%\), \(69.1 \%\) respectively.  The percolation are quantitatively described in Eq.\eqref{percolation}.}
    \label{figpercolation}
\end{figure}

To quantitatively characterize which fraction of particles (\(n_{\text{particle}}\)) contributes a given fraction (\(n_{\text{DOS}}\)), we perform a percolation analysis to select the most dominant particles, as shown in Fig.\ref{figpercolation}. The top \(n_{\text{particle}} = 9.2\%\), \(24.5\%\), \(44.6\%\), \(69.1\%\) of particles contribute \(n_{\text{DOS}} = 20\%\), \(40\%\), \(60\%\), \(80\%\) of the total DOS around the BP frequency. The fraction of particles \(n_{\text{particle}}\) that contributes a given fraction \(n_{\text{DOS}}\) of the total DOS can be determined from the relation
\begin{equation}
\begin{split}
    n_{\text{DOS}} = &\dfrac{1}{\int_{0}^{0.2} g(\omega) d\omega} \times \\ & \sum_{a = 1}^{N n_{\text{particle}}} \mathcal{RO}_a \left[ \int_{0}^{0.2} \left( g_{a,x}(\omega) + g_{a,y}(\omega) \right) d\omega \right], \label{percolation}
\end{split}
\end{equation}
once \(n_{\text{DOS}}\) is fixed. Here the integration interval \(\omega \in [0,0.2]\) corresponds to the BP region. The operator \(\mathcal{RO}_a [\cdot]\) denotes reordering the atom indices \(a\) according to \(\int_{0}^{0.2} (g_{a,x}(\omega) + g_{a,y}(\omega)) d\omega\) from largest to smallest; \(g_{a,x(y)}(\omega)\) is the local DOS of the \(x(y)\) degree of freedom of atom \(a\). After reordering, we have
\begin{equation}
\begin{split}
    &\int_{0}^{0.2} (g_{1,x}(\omega) + g_{1,y}(\omega)) d\omega \\
    > & \int_{0}^{0.2} (g_{2,x}(\omega) + g_{2,y}(\omega)) d\omega \\
    > & \dots \\
    > & \int_{0}^{0.2} (g_{N,x}(\omega) + g_{N,y}(\omega)) d\omega . \label{ordering}
\end{split}
\end{equation}
Morphologically, the contributing particles are rather uniformly distributed in space, whether they form small clusters under extreme percolation thresholds\cite{Hu2022,PhysRevLett.133.188302}, or a connected complex pattern under more inclusive thresholds\cite{ch39-6bhs}. In this situation, it is not justified to claim that the low-frequency total DOS is dominated by a minority of particles.

The above results confirm the presence of vibrational inhomogeneity in our simulated two-dimensional amorphous solid. Although the frequency-resolved vibrational inhomogeneity (the local DOS) determines the total DOS features such as the BP and the Van Hove peak, its spatial distribution does not exhibit clear signatures that distinguish different regions of the total DOS. Consider the crystal case in Fig.\ref{figdosandpr}: from a dynamical or total DOS perspective, the Van Hove peak and the low-frequency region are distinct because the latter corresponds to sound waves while the former does not. This distinction cannot be captured by the distribution of dynamical response intensity, which is perfectly homogeneous at all frequencies in a crystal. Similarly, the BP in the amorphous solid is accompanied by only a slight dip in the frequency-resolved participation ratio (Fig.\ref{figdosandpr}). In our view, this correlation is too weak to imply causation, given the sensitivity of the participation ratio discussed earlier. Moreover, Figs.\ref{figldosnew} and \ref{figldosmore} show that the participation ratio is largely determined by a small minority of particles that are too few to affect the total DOS. These results suggest that one should reconsider whether local vibrational inhomogeneity provides a satisfactory route to explain global dynamical anomalies.

In the introduction, we mentioned several dynamical anomalies such as localized modes\cite{Wang2019,PhysRevLett.108.095501,10.1063/5.0147889}, soft spots\cite{Corrado2020}, dynamical defects\cite{ch39-6bhs,PhysRevResearch.6.023053}, and dynamical excitations\cite{Hu2022,PhysRevLett.133.188302}. Most of these anomalies are identified through eigenmode analysis and require separating phonon-like and non-phononic contributions. These anomalies typically involve a minority of particles that behave differently from the rest. The present work does not question the existence of such dynamical anomalies (vibrational inhomogeneity); however, their contribution to total DOS features like the BP should be carefully considered, because these features actually arise from the majority of particles. 

\section{Conclusion}
In summary, we use Green's function method combined with simulation data to better probe the spatial distribution of the dynamical response in amorphous solids. The order parameter, frequency-resolved participation ratio, is proposed to character the level of vibrational inhomogeneity for a specific frequency. Vibrational inhomogeneity is confirmed and visualized through three complementary approaches: spatial maps, statistical distributions, and percolation analysis over given frequency ranges. The results indicate that while vibrational inhomogeneity is dominated by a minority of particles, global dynamical anomalies such as BP are governed by the vast majority of particles. These findings provide an alternative perspective on the role of vibrational inhomogeneity in amorphous solids.

\section*{Acknowledgments}
The author would like to thank Qing Xi for performing the simulation and offering important discussions throughout this work.

\section*{Data availability}
There are no publicly available research data or software supporting this manuscript. Requests for further information or data should be sent to the authors.

\section*{Appendix A: Effect of the parameter \(\eta\) in the Green's function}
\begin{figure}[htbp]
    \centering
    \includegraphics[width=0.8\linewidth]{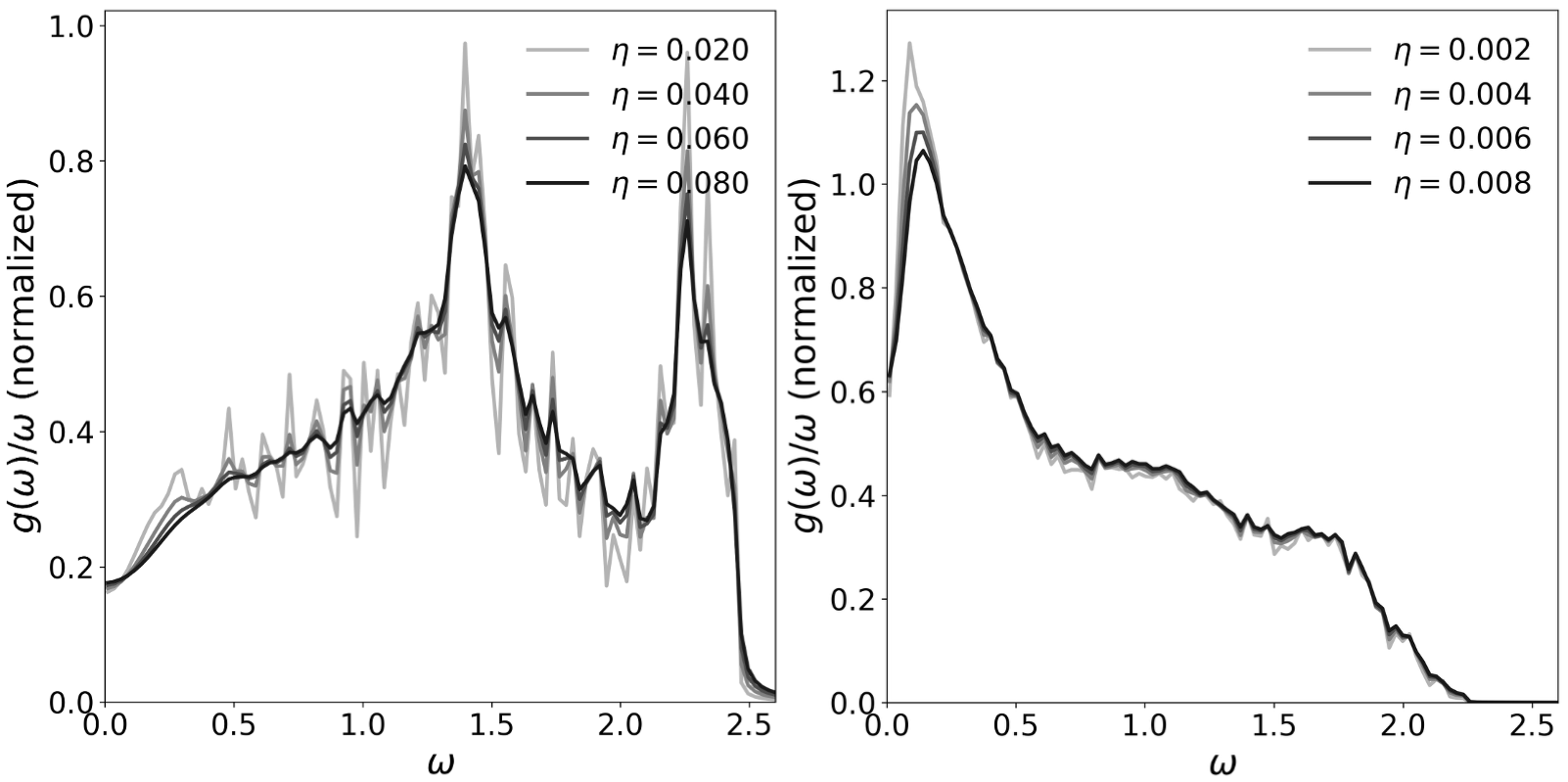}
    \caption{Reduced DOS for the crystal (\textbf{left}) and amorphous solid (\textbf{right}) computed via the Green's function method [Eq.\eqref{dos}] with different values of the broadening parameter \(\eta\). A reasonably larger \(\eta\) helps to smooth the curve.}
    \label{eta}
\end{figure}

In the definition of the Green's function,
\begin{equation}
    G(\omega) = [(\omega^2 + i\eta) I - H]^{-1},
\end{equation}
the parameter \(\eta \to 0^+\) introduces an imaginary part, which allows one to apply the Plemelj identity (see Ref.\cite{PhysRevLett.122.145501})
\begin{equation}
    \lim_{\eta \to 0^+} \dfrac{1}{\pi} \text{Im}\left[\dfrac{1}{x + i \eta} \right] = -\delta(x),
\end{equation}
to compute the total DOS as
\begin{equation}
    g(\omega) = -\omega \mathrm{Im}[\mathrm{Tr}[G(\omega)]] = \sum_\alpha \delta (\omega - \omega_\alpha).
\end{equation}
A smaller \(\eta\) makes the delta function narrower and closer to a true delta function, whereas a larger \(\eta\) broadens the delta function and yields a smoother DOS curve, albeit at the expense of the strict delta-function representation. Figure \ref{eta} shows the DOS results for several values of \(\eta\). The DOS is robust to small variations in \(\eta\). The choice of \(\eta\) thus balances fidelity to the true spectrum and curve smoothness.

\end{document}